\documentclass[apj]{emulateapj}
\usepackage{apjfonts}
\usepackage{amssymb}
\usepackage{natbib}
\usepackage{graphicx}
\usepackage{color}
\usepackage[dvipdfmx]{hyperref}

\def\mbh{{M_{\bullet}}}

\def\rd{{d}}
\def\lsph{{L_{\rm sph}}}
\def\msph{{M_{\rm sph}}}

\def\lax{{$\mathrel{\hbox{\rlap{\hbox{\lower4pt\hbox{$\sim$}}}\hbox{$<$}}}$}}
\def\gax{{$\mathrel{\hbox{\rlap{\hbox{\lower4pt\hbox{$\sim$}}}\hbox{$>$}}}$}}

\slugcomment{To appear in the Astrophysical Journal}
\shorttitle{COSMOLOGICAL EVOLUTION OF SMBHS. I.}
\shortauthors{LI, HO, \& WANG}

\begin{document}

\title{Cosmological evolution of supermassive black holes. I. mass function at $0<z\lesssim2$}

\author{Yan-Rong Li\altaffilmark{1}, Luis C. Ho\altaffilmark{2}, and 
        Jian-Min Wang\altaffilmark{1, 3}}

\altaffiltext{1}
{Key Laboratory for Particle Astrophysics, Institute of High Energy Physics,
Chinese Academy of Sciences, 19B Yuquan Road, Beijing 100049, China; 
liyanrong@mail.ihep.ac.cn, wangjm@mail.ihep.ac.cn}

\altaffiltext{2}
{The Observatories of the Carnegie Institution for Science, 813 Santa Barbara Street, Pasadena, CA 91101, USA; lho@obs.carnegiescience.edu}

\altaffiltext{3}
{National Astronomical Observatories of China, Chinese Academy of Sciences, 20A Datun Road, 
Beijing 100020, China}

\begin{abstract}
We present the mass function of supermassive black holes (SMBHs) over the 
redshift range $z=0-2$, using the latest deep luminosity and mass functions 
of field galaxies to constrain the masses of their spheroids, which we relate 
to SMBH mass through the empirical correlation between SMBH and spheroid mass 
(the $\mbh-\msph$ relation).  In addition to luminosity fading of the stellar 
content of the spheroids, we carefully consider the variation of the 
bulge-to-total luminosity ratio of the galaxy populations and the $\mbh/\msph$ ratio, 
which, according to numerous recent studies, evolves rapidly with redshift.  
The SMBH mass functions derived from the galaxy luminosity and mass functions 
show very good agreement, both in shape and in normalization.  The resultant 
SMBH mass function and integrated mass density for the local epoch ($z \approx 
0$) match well those derived independently by other studies.  Consistent with 
other evidence for cosmic downsizing, the upper end of the mass function 
remains roughly constant since $z\approx2$, while the space density of lower 
mass black holes undergoes strong evolution.  We carefully assess the impact 
of various sources of uncertainties on our calculations.  A companion paper 
uses the mass function derived in this work to determine the radiative 
efficiency of black hole accretion and constraints that 
can be imposed on the cosmological evolution of black hole spin.
\end{abstract}


\section{Introduction}
Supermassive black holes (SMBHs) are ubiquitous in galaxies with a central 
stellar bulge (e.g., \citealt{Magorrian98, Ho99, Kormendy04}), and, via the 
process of mass accretion (e.g., \citealt{sal64, Zeldovich64, Lynden_Bell69}), 
power active galactic nuclei (AGNs).  The discovery of strong correlations 
between SMBH mass and the overall properties of the host spheroid, in 
particular the stellar luminosity, mass, and velocity dispersion 
(\citealt{Kormendy95, Magorrian98, Gebhardt00, Ferrarese00, Haring04}), 
has generated intense interest in the notion that black hole and host galaxy 
growth are closely connected.  Studying the mass function of SMBHs and its 
evolution with redshift is therefore of great significance not only for 
understanding their cosmological evolution of mass and spin 
(\citealt{Shapiro05, Wang06, Wang08, Wang09}), but also for addressing many issues in galaxy 
formation (e.g., \citealt{Di_Matteo05, Shankar09}). 

Generally speaking, there are two approaches to investigate the 
SMBH mass function (e.g., \citealt{Tamura06, Shankar09}). The first approach, 
which we refer to as theoretically based, stems from the pioneering work of 
\cite{Soltan82}. The standard procedure, based on the argument that black hole 
growth is mainly driven by mass accretion (e.g., \citealt{Small92, Marconi04, 
Cao08, Shankar_et09b,Cao10}), is to integrate the continuity equation to 
determine the number density of SMBHs and to couple it to the AGN luminosity 
function (LF).  Such calculations usually assume two parameters: the radiative 
efficiency for energy conversion of accretion and the Eddington ratio of AGNs. 
As a result, comparing between the predicted SMBH mass function at $z=0$ with
the locally observed quantity sets interesting constraints on the radiative 
efficiency.  The radiative efficiency, averaged over redshift and black hole 
mass, is found to be $\eta\approx 0.1$ (\citealt{Yu02, Marconi04}).

The observational approach uses empirical scaling relationships between SMBH 
mass and host properties, in concert with the LF or 
stellar mass function (SMF) of the host galaxies, to infer statistical 
information on the SMBH population (see \citealt{Shankar09} for a review). 
Most previous works, because of the availability of ample data, have focused 
on the local ($z \approx 0$) SMBH mass function (e.g., \citealt{Graham_et07, 
Vika09, Shankar_et09b}).  Few have treated the evolution of the SMBH mass 
function. \cite{Shankar_et09a} predicted SMBH mass function at $0<z<6$ using 
the local velocity dispersion function of spheroids together with a model for 
their age distribution; however, their results are highly dependent on the
assumed age distribution of the spheroids, which is poorly constrained.  Out to 
intermediate redshifts ($z\approx 1$), \cite{Tamura06} determined the SMBH 
mass function by employing a non-evolving relation between SMBH mass and 
spheroid luminosity ($\mbh-\lsph$ relation) and by assuming that spheroids 
evolve passively.   A number of recent studies, however, suggest that the
black hole-host scaling relations evolve with redshift (\citealt{McLure06, 
Peng06a, Peng06b, Ho07, Bennert10, Bennert11, Decarli10, Merloni10}). Although 
the exact magnitude of the effect is not yet well-defined, most studies 
find that the ratio of black hole mass to spheroid mass is higher in the past.
It is essential to take into account these evolutionary effects. At the same 
time, outstanding progress has been made on deep multiwavelength surveys of 
normal galaxies and AGNs (e.g., \citealt{Lawrence07, Abazajian09}). This 
allows us to construct SMBH mass functions more reliably and to extend them 
to even higher redshifts than was possible before (e.g., \citealt{Tamura06, 
Vika09}, and references therein). 

In this work, we aim to derive the SMBH mass function using an up-to-date 
galaxy LF and SMF, with careful inclusion of various redshift-dependent 
effects. We extend and update the work of \cite{Tamura06} out to $z\approx2$. 
It is interesting to emphasize that once the SMBH mass function is obtained, 
a combination of the above two approaches would immediately inform us of the 
redshift evolution of the radiative efficiency, from which inferences on 
black hole spins can be made.  This is the subject of a companion paper. 

This paper is organized as follows. In Section 2, we describe the procedure 
for deriving the SMBH mass function. Section 3 presents the resultant mass 
functions and performs a comparison with previous works. We then investigate 
the uncertainties in our calculations and show how they impact upon the 
results (Section 4). We discuss the results of this work and the conclusions
in Section 5.

\section{Derivation of SMBH mass functions}
We employ two methods to derive the SMBH mass function: galaxy LFs, and, 
alternatively, galaxy SMFs.

We use the recent $K$-band galaxy LF out to $z\approx4$ obtained by 
\citet{Cirasiolo10}, who utilized the first data release of 
the UKIDSS Ultra Deep Survey. In addition to its depth and large survey area,
the Ultra Deep Survey provides reliable photometry in 16 bands.   
The $K$ band is ideal for this application because it is not severely affected 
by dust absorption and because it effectively traces the old 
stellar population and hence is a reliable indicator of stellar mass.

\citet{Perez-Gonzalez2008} computed SMF for a sample of $0<z<4$ galaxies for 
which they could obtain rest-frame near-infrared photometry using 
{\it Spitzer}/Infrared Array Camera observations.  They calculated stellar masses and 
photometric redshifts by fitting the observational data to $\sim$2000 
reference templates for which there are reliable spectroscopic redshifts and 
well-covered spectral energy distributions from the ultraviolet to the 
mid-infrared bands.  The stellar emission models were computed using the code
PEGASE (\citealt{Fioc97}), adopting a Salpeter (1955) initial mass function 
(IMF) with stellar mass between $0.1~M_\odot$ and $100~M_\odot$.

To derive the SMBH mass function from the galaxy LF or SMF, we need three 
ingredients: (1) the bulge-to-total luminosity ratio (B/T)  to transform the 
total galaxy luminosity to the spheroid luminosity, for which we assume that 
the light distribution follows the mass distribution in galaxies;
(2) a scaling relation between SMBH mass and the mass of the bulge of the host;
and, for the method involving the galaxy LF, (3) a prescription to describe the 
passive evolution of the spheroid luminosity to account for the evolution of 
the stellar population. 
Here we implicitly assume that all the galaxies under consideration possess a 
bulge component that hosts a black hole (\citealt{Ho08}), that the black hole 
mass scales with the bulge mass, and that the scaling relation evolves with 
redshift, as described below.

%
\begin{figure}[t]
\centering
\includegraphics[angle=-90.0, width=0.40\textwidth]{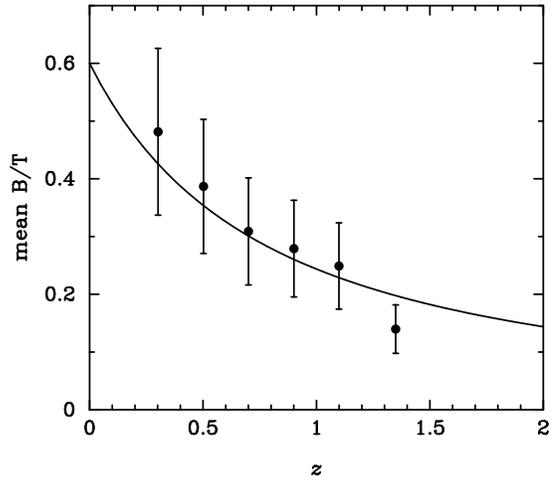}
\caption{Evolution of the bulge-to-total luminosity ratio, B/T. Data points 
show the mean B/T based on the type-dependent galaxy LFs of \cite{Zucca06}, 
and the error bars follow from the galaxy LFs. The solid line represents B/T = 
$0.6(1+z)^{-1.3}$ (see the text for details).}
\label{bt_fig}
\end{figure}
%
%
%
\begin{figure}[t]
\centering
\includegraphics[angle=-90.0, width=0.40\textwidth]{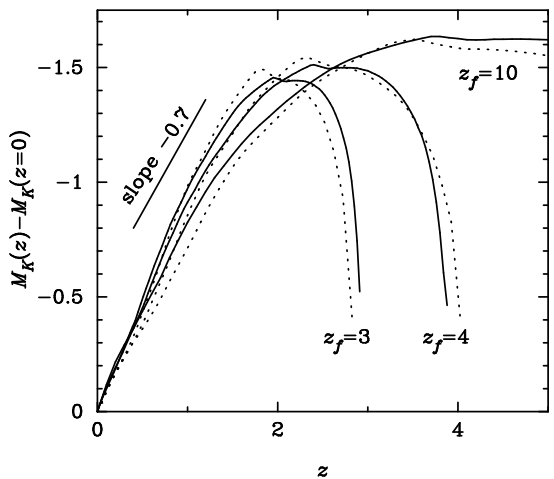}
\caption{Passive evolution of the rest-frame $K$-band absolute magnitude, 
calculated using PEGASE (solid lines) and GALAXEV (dotted lines), with 
formation redshifts $z_f$ = 3, 4, and 10.  For $z$ \lax\ 1.5 the curves are 
well approximated by a straight line with a slope of $\sim -0.7$.}
\label{passive_fig}
\end{figure}

Throughout the paper, we adopt a cosmological model with $\Omega_{\rm m}=0.3$, 
$\Omega_\Lambda=0.7$, and $H_0=70\,{\rm km~s^{-1}~Mpc^{-1}}$. Unless otherwise 
specified, magnitudes are given in the AB system (\citealt{Oke83}). According 
to \cite{Hewett06}, for the UKIDSS photometric system, the flux density for the
zero point of the Vega-based magnitudes is 
$f_\nu=6.31\times10^{-21}$ erg s$^{-1}$ cm$^{-2}$ Hz$^{-1}$.
The magnitude offset between the Vega-based system and the AB system is 
$K_{\rm AB}=K_{\rm Vega}+1.9$ (\citealt{Hewett06}). We adopt a solar $K$-band 
luminosity of $L_{\odot, K}=0.82\times10^{32}\,{\rm erg~s^{-1}}$ (see Table 2.1 
in \citealt{Binney98}), which corresponds to a solar $K$-band absolute 
magnitude of $M_{\odot, K}=5.2$ in the AB system.
%
%
%
\begin{figure}[t]
\centering
\includegraphics[angle=-90.0, width=0.40\textwidth]{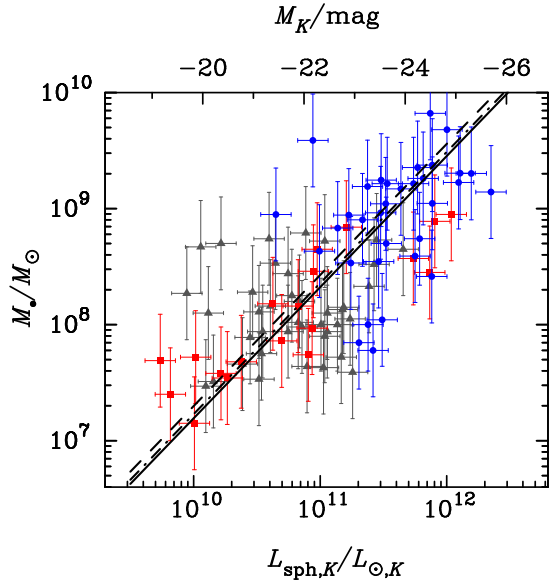}
\caption{$\mbh-L_{\rm sph}$ relation in the $K$ band.  Lines show the 
$\mbh-L_{\rm sph}$ relation described by Equation (\ref{mlz_equ}) at $z=0$ 
(solid), $z=1$ (dashed), and $z=2$ (dot-dashed). The values of the free 
parameters are listed in Table 1. Superimposed for comparison are the 
observational data from Table 3 of Peng et al. (2006b; blue points) and Tables 
2 and 3 of Bennert et al. (2010; black triangles and red squares).  Note that the 
tabulated data from \cite{Peng06b} and \cite{Bennert10} were not corrected for 
luminosity evolution; the term accounting for the evolution of the spheroid 
luminosity in Equation (\ref{mlz_equ}) is essential to perform a direct 
comparison. We adopt $k$-corrections appropriate for early-type galaxies, 
assuming $K-V=-2.79$ mag and $K-R=-2.18$ mag (\citealt{Fukugita95, Girardi03}).}
\label{ml_fig}
\end{figure}
%
%
\subsection{Bulge-to-total Luminosity Ratio}
B/T is an important morphological classification parameter that gives the 
fraction of the total luminosity contained in the bulge.  Observations of 
local early-type galaxies show that they possess high values of B/T, while 
late-type galaxies possess lower values of B/T (e.g., \citealt{Simien86, 
Schechter87}). Many studies have performed bulge-to-disk decomposition of 
images of galaxies to quantitatively estimate their B/T (e.g., 
\citealt{Kormendy77, Kent85, Allen06, Benson07, McGee08, Tasca11}). Although 
its distribution is broad, B/T generally increases with galaxy luminosity 
(\citealt{Schechter87, Benson07, Tasca11}). 
Below we will show how to determine the average value of B/T at $z=0$, motivated
by observations and the comparison between our calculated SMBH mass function and 
the local observed one.

It is difficult to constrain B/T at high redshifts because of the challenges
involved in obtaining reliable bulge-to-disk decompositions for distant 
galaxies. However, we can use the fact that certain characteristic properties 
of galaxies statistically correlate with B/T to estimate it indirectly at high 
redshifts. \cite{Zucca06}, using template galaxy spectral energy distributions, 
classified galaxies out to $z\approx1.5$ into four spectral types---ranging 
from early-type to irregulars---and compiled morphological type-dependent 
galaxy LFs. Their four spectral types roughly correspond to the morphological 
types E/S0, Sa--Sb, Sc--Sd, and Irr. Based on the average 
properties of nearby galaxies (e.g., \citealt{Kent85, Weinzirl09, Tasca11}), 
we assign B/T $\approx$ 0.7, 0.3, 0.1, and 0.0, respectively, to the above 
four type bins.  Figure \ref{bt_fig} plots the resultant variation of B/T with 
redshift, where the error bars follow from those of the galaxy LFs.  The 
trend is physically reasonable: the mean B/T decreases systematically and 
monotonically with increasing redshift.  The evolutionary trend can be 
parameterized as (see also \citealt{Merloni04})

\begin{equation}
{\rm B/T} = 0.6(1+z)^{-\gamma},
\label{bt_equ}
\end{equation}
with $\gamma>0$, where the normalization is set to B/T = 0.6 at $z = 0$.  As 
shown in Figure \ref{bt_fig}, $\gamma = 1.3$ gives a reasonably good 
description of the available data.

This evolutionary trend of B/T can be understood within the framework of 
hierarchical galaxy formation (e.g., \citealt{Croton06, Khochfar06, 
Weinzirl09}), wherein spheroids are built up through two physical processes, 
mergers and secular evolution.  Whereas major mergers scramble disks into
(classical) bulges, multiple successive minor mergers can also transform a 
spiral galaxy into an elliptical (\citealt{Bournaud07}). Secular processes
from global, non-axisymmetric structures (e.g., bars and spiral arms) in 
isolated galaxies further contribute to (pseudo-)bulge growth 
(\citealt{Kormendy04}, and references therein). 

%
\begin{figure*}[t]
\centering
\includegraphics[angle=-90.0, width=0.9\textwidth]{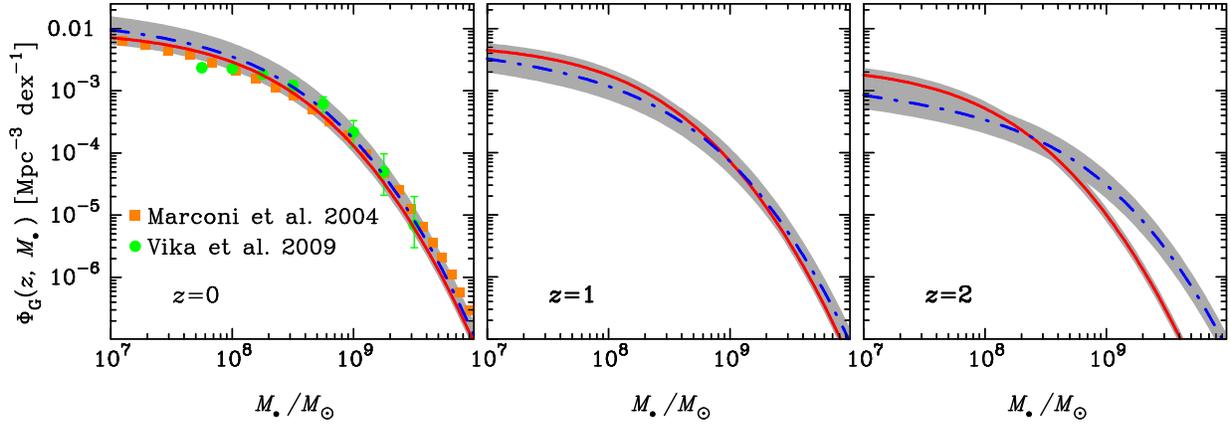}
\caption{SMBH mass function at $z$ = 0, 1, and 2, derived from the galaxy LF 
(red solid lines) and the galaxy SMF (blue dot-dashed lines).  The values of 
the free parameters are listed in Table 1.  Shaded areas represent the errors 
from the galaxy LF and SMF. In the $z=0$ panel, the orange squares mark the 
local SMBH mass function from \citet{Marconi04}, and the green solid points 
give the corresponding derivation from \citet{Vika09}, whose mass limit is 
$\mbh \approx 10^{7.7}\,M_\odot$.}
\label{nbh_fig}
\end{figure*}
%
%
%
\begin{deluxetable*}{ccl}
\tabletypesize{\footnotesize}
\tablecolumns{3}
\tablewidth{0pc}
\tablecaption{Free parameters}
\tablehead{
\colhead{~~~~Parameter~~~~} & \colhead{~~~~~~Value~~~~~~}    & \colhead{~~~~~~Implication~~~~~~}}
\startdata
B/T      & 0.6 & Ratio of bulge-to-total luminosity (B/T) at $z=0$\\
$\gamma$ & 1.3 & Power-law index for the redshift evolution of B/T (Equation \ref{bt_equ})\\
$\beta$  & 1.4 & Power-law index for the redshift evolution of $\frac{\mbh}{\msph}$ (Equation \ref{mme_equ})\\
$Q$      & 0.7 & Passive evolution of $K$-band magnitude (Equation \ref{passive_equ})
\enddata
\end{deluxetable*}
%
%
%
\begin{figure}[t]
\centering
\includegraphics[angle=-90.0, width=0.4\textwidth]{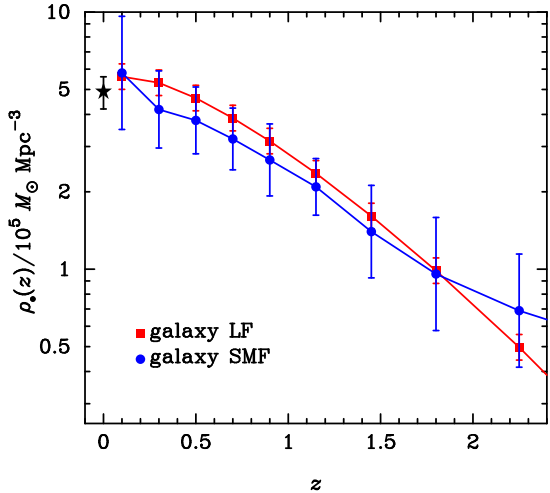}
\caption{SMBH mass density ($\rho_\bullet$) as a function of redshift for 
$\mbh > 10^7\,M_\odot$ 
derived from the galaxy LF (red squares) and the galaxy SMF (blue points). 
Redshift bins correspond to the bins chosen by \cite{Perez-Gonzalez2008}. The 
black star symbol shows the locally observed value of $\rho_\bullet=
(4.9\pm0.7)\times10^5 \,M_\odot \,{\rm Mpc^{-3}}$ \citep{Vika09}.}
\label{nbhz_fig}
\end{figure}
%
%
%
\begin{figure}[t]
\centering
\vspace*{0.1cm}
\includegraphics[angle=-90.0, width=0.41\textwidth]{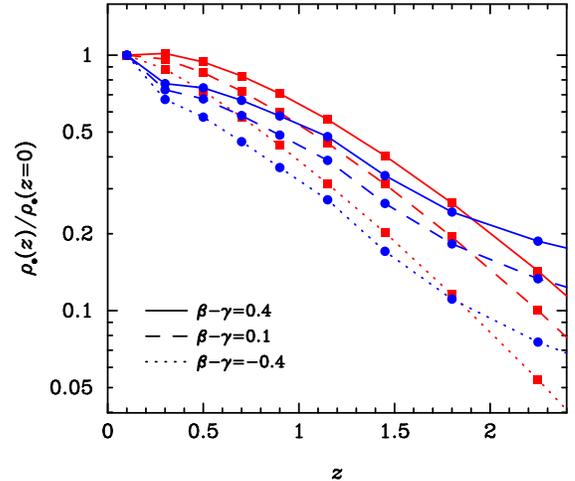}
\caption{SMBH mass density ($\rho_\bullet$) as a function of redshift, 
normalized to $z = 0$, for $\mbh > 10^7\,M_\odot$ and for 
different combinations of $\beta-\gamma$, derived from the galaxy LF
(red squares) and the galaxy SMF (blue points). Redshift bins 
correspond to the bins chosen by \cite{Perez-Gonzalez2008}.}
\label{beta_fig}
\end{figure}
%
%

%
\subsection{The $\mbh-\msph$ and $\mbh-\lsph$ Relation}
We parameterize the relations between black hole mass and spheroid mass 
and $K$-band luminosity as
\begin{equation}
\log\frac{\mbh}{M_\odot}=a_M\log\frac{\msph}{M_\odot} + b_M,
\end{equation}
and
\begin{equation}
\log\frac{\mbh}{M_\odot}=a_L\log\frac{L_{{\rm sph},K}}{L_{\odot, K}} + b_L,
\label{ml_equ}
\end{equation}
where $(a_M, b_M, \Delta_M)$ and $(a_L, b_L, \Delta_L)$ are parameters to be 
determined from the data, with $\Delta_M$ and $\Delta_L$ being the 
respective intrinsic scatter of each relation.  In our calculations, we adopt 
the $\mbh-\msph$ relation derived by \cite{Haring04} and the $\mbh-\lsph$ 
relation derived by \cite{Marconi03}, wherein $(a_M, b_M, \Delta_M)=
(1.12\pm0.06, -4.12\pm0.10, 0.30)$ and $(a_L, b_L, \Delta_L)=(1.13\pm0.12, 
-4.11\pm0.07, 0.31)$. 

%
\begin{figure*}[t]
\centering
\includegraphics[angle=-90.0, width=0.9\textwidth]{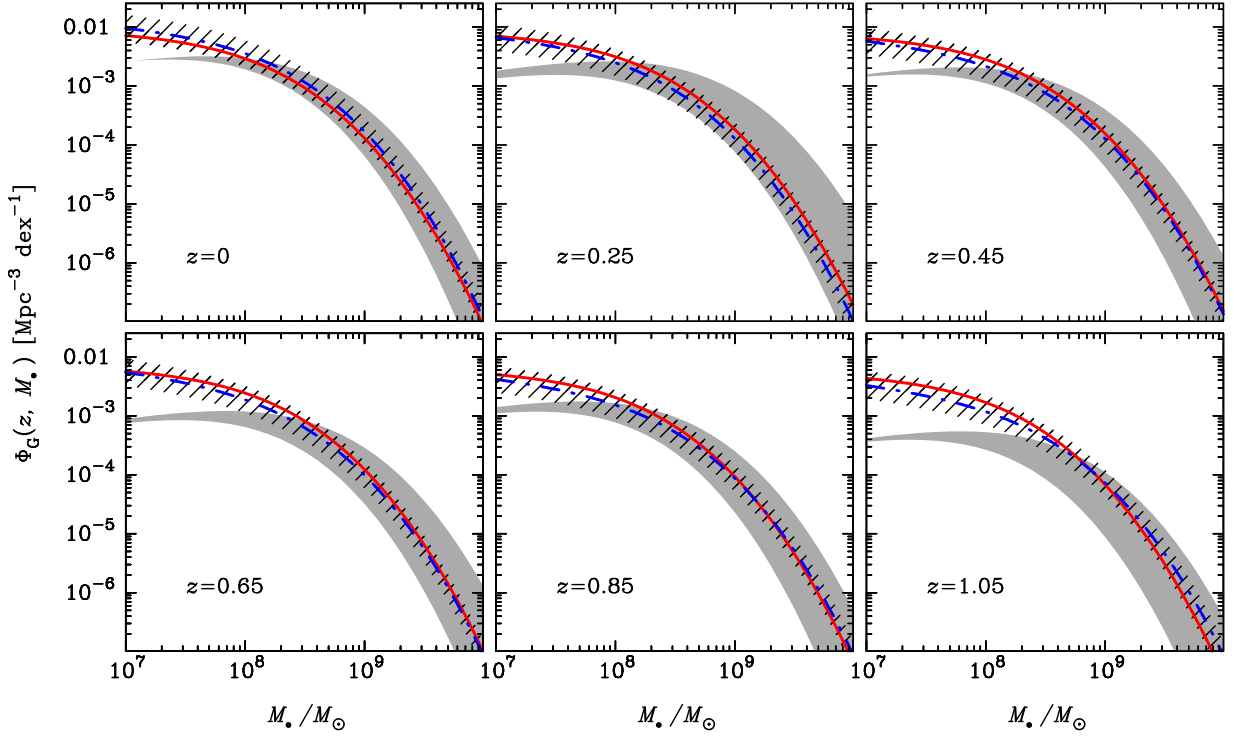}
\caption{Comparison of our SMBH mass function with that derived by 
\cite{Tamura06} for different redshift bins. Solid lines and dot-dashed lines 
are the mass functions derived from the galaxy LF and galaxy SMF, 
respectively, with hatched areas showing the errors. The shaded area denotes 
the mass functions derived by \cite{Tamura06}.}
\label{tamura_fig}
\end{figure*}

We emphasize that these two correlations pertain solely to nearby galaxies.
Redshift-dependent effects should be considered when applying them to higher 
redshifts. As mentioned above, mounting evidence suggests that the scaling 
relations between black hole mass and host galaxy evolve with redshift
(\citealt{McLure06, Peng06a, Peng06b, Ho07, Bennert10, Bennert11, Decarli10, 
Merloni10}). Broadly speaking, the ratio of the central SMBH mass and the 
stellar bulge mass, $\mbh/\msph$, tends to be higher at higher redshifts. If we
express this redshift dependence as
\begin{equation}
\frac{\mbh}{\msph}\propto(1+z)^\beta,
\label{mme_equ}
\end{equation}
with $\beta>0$, the $\mbh-\msph$ relation at any given redshift $z$ can be 
rewritten
\begin{equation}
\log\frac{\mbh}{M_\odot}=a_M\log\frac{\msph}{M_\odot} + b_M + \beta\log(1+z). 
\end{equation}
Similarly, the $\mbh-\lsph$ relation needs to be modified with the additional
term $\beta\log(1+z)$.

Although most recent studies seem to be converging on the idea that black hole 
growth precedes bulge assembly at higher redshifts, there is no uniform 
consensus on the exact magnitude of the effect (i.e., the value of $\beta$).
\cite{McLure06} estimated the black hole-to-spheroid mass ratio in a sample of 
radio-loud AGNs, hosted by massive, early-type galaxies in the redshift range 
$0<z<2$, and found that $\beta\approx2$. \cite{Peng06b}, analyzing the 
observed $R$-band $\mbh-\lsph$ relation for a sample of 31 gravitationally 
lensed AGNs and 20 unlensed AGNs at $1\lesssim z \lesssim 4.5$, concluded that,
after accounting for luminosity evolution, the $\mbh/\msph$ at $z>1.7$ is
$\sim$4 times larger than the local value.  The implied value of $\beta$ is 
$\sim 1.4$.  In a study of 89 broad-line AGNs between $1<z<2.2$ selected from 
the zCOSMOS survey, \cite{Merloni10} inferred a somewhat weaker evolution for 
the ratio of black hole mass to host galaxy stellar mass, with 
$\beta\approx0.68$; these authors did not explicitly separate the spheroid 
mass from the total galaxy mass, and it is possible that $\mbh/\msph$ itself 
may evolve faster than indicated.  A relatively strong degree of evolution in 
$\mbh/\msph$, consistent with $\beta\approx 1.3-1.4$, was reported by 
\cite{Bennert10} for $z \approx 0.4$ and 0.6, by \cite{Decarli10} for $0<z<3$, 
and by \cite{Bennert11} for $1<z<2$.

Based on the above summary, the weight of the current evidence suggests 
that $\mbh/\msph$ evolves with redshift, and, for the parameterization 
given in Equation (\ref{mme_equ}), that a reasonable choice for the redshift dependence 
of the evolution is $\beta=1.4$.  We choose this as a fiducial value. Below we 
will show that the physical requirement that black hole mass density 
increases monotonically over time places additional constraints on $\beta$.

%
\subsection{Passive Evolution of the Spheroid Luminosity}
When considering the $\mbh-\lsph$ relation at high redshift---a necessary 
step for converting the galaxy LF into the SMBH mass function---a correction to 
the spheroid luminosity must be applied to account for the fading of the 
stellar population with time.  (Note that our alternative procedure of using 
the galaxy SMF bypasses this complication, and thus serves as a useful 
consistency check for our LF-based method.) In practice, the correction 
depends on the star formation history of the spheroid.  Following common 
practice, we employ a single-burst star formation model characterized by an 
$e$-folding time $\tau$ and a formation redshift $z_f$.  Such a passive 
evolution scenario traces a single, maximally old stellar population and gives 
a conservative estimate of the amount of luminosity evolution.

We run the code PEGASE (version 2.0; \citealt{Fioc97}) and GALAXEV 
(\citealt{Bruzual03}) to compute the amount of luminosity evolution. Both sets
of models use stellar evolutionary tracks from the Padova 1994 library. We 
assume a Salpeter IMF, a lower mass limit of 0.1 $M_\odot$, an upper mass 
limit of 100 $M_\odot$, and solar metallicity.  The $e$-folding time is set 
nominally to $\tau=1$ Gyr, and the formation redshift is set to $z_f=3, 4,$ 
and 10. The evolution of the rest-frame $K$-band luminosity is shown in 
Figure \ref{passive_fig}.  We find that the results are quite insensitive to 
$z_f$.   
For completeness, we also explore models assuming a 
Chabrier (2003) IMF, as well as metallicities ranging from 0.4 $Z_\odot$ 
to 2.5 $Z_\odot$, and $\tau=0.5-2$ Gyr.  The results are similar.
In all cases, for $z\lesssim2 $, the 
luminosity evolution in the $K$ band can be described as
\begin{equation}
 M_K(z)=M_K(z=0)-Qz,
 \label{passive_equ}
\end{equation}
where $Q\approx0.7$. Including this prescription for luminosity evolution,
we can write the $\mbh-\lsph$ relation at redshift $z$ as
\begin{equation}
\log\frac{\mbh}{M_\odot}=a_L\left(\log\frac{\lsph}{L_\odot} -
\frac{Qz}{2.5}\right) + b_L + \beta\log(1+z). 
\label{mlz_equ}
\end{equation}

Figure \ref{ml_fig} plots the $\mbh-L_{\rm sph}$ relation for the set of free 
parameters tabulated in Table 1. The observed data from Table 3 of 
\cite{Peng06b} and Tables 2 and 3 of \cite{Bennert10} are superimposed for 
comparison. (Note that the original data were not corrected for luminosity evolution.) 
This comparison illustrates an important fact: a single $\mbh-\lsph$ relation 
adequately describes the data from $z=0$ to $z=2$ {\it after}\ applying 
our prescriptions for the redshift dependence of $\mbh/\msph$ and 
spheroid luminosity evolution.  This verifies the robustness of our 
redshift-dependent corrections and reaffirms the conclusions of previous 
studies, such as those of \cite{Peng06a, Peng06b}, \cite{Bennert10, Bennert11},
and \cite{Decarli10}.

Having introduced the above three components (for the redshift evolution of 
B/T, $\mbh/\msph$, and bulge luminosity), we can relate the SMBH mass function 
to the galaxy LF, accounting for intrinsic scatter in the 
$\mbh-\lsph$ relation:
\begin{eqnarray}\nonumber
\Phi_{\rm G}(z,\mbh)&=&2.5\frac{\rd\log L_{{\rm sph},K}}{\rd \mbh}
\int\rd\log L_{{\rm sph},K}\Phi(z, M_K)\\
& &\times\frac{1}{\sqrt{2\pi}\Delta_L}\exp
\left[-\frac{(\log L_{{\rm sph},K}-\log\mbh)^2}{2\Delta_L^2}\right],
\label{nbh_equ}
\end{eqnarray}
where $M_K$ is the total $K$-band absolute magnitude of the galaxy, 
$L_{{\rm sph},K}$ is the luminosity of the bulge, $\Phi(z, M_K)$ is the 
$K$-band galaxy LF given by Cirasuolo et~al. (2010), and $\Phi_{\rm G}$ 
measures the mass function per unit volume per unit mass. Here the subscript 
``G'' distinguishes the mass function of galaxies from that of AGNs. 
The factor $\rd\log L_{{\rm sph},K}/\rd\mbh=1/[\ln(10)a_L\mbh]$, according to 
Equation (3). The transformation between $L_{{\rm sph},K}$ and $M_K$ reads
\begin{equation}
M_K= M_{\odot, K} 
-2.5 \log \left(\frac{1}{\rm B/T}\frac{L_{{\rm sph},K}}{ L_{\odot, K}}\right).
\label{mag_equ}
\end{equation}
Expressing the galaxy LF as a Schechter (1976) function (Cirasuolo et~al. 2010),
\begin{equation}
\Phi(z, M_K)=0.4\ln(10)\Phi_010^{-0.4(\alpha+1)\Delta M}\exp\left[-10^{-0.4\Delta M}\right],
\end{equation}
with $\Delta M=M_K-M_K^*$,
\begin{equation}
 M_K^*(z)=M_K^*(z=0)-\left(\frac{z}{z_M}\right)^{k_M},
\label{mk_equ}
\end{equation}
and
\begin{equation}
\Phi_0(z)=\Phi_0(z=0)\exp\left[-\left(\frac{z}{z_\phi}\right)^{k_\phi}\right].
\label{phi0_equ}
\end{equation}
The free parameters are $\alpha=-1.07\pm0.1$, $z_M=1.78\pm0.15$,
$k_M=0.47\pm0.2$, $z_\phi=1.70\pm0.09$, $k_\phi=1.47\pm0.1$,
$M_K^*(z=0)=-22.26$ (fixed), and $\Phi_0(z=0)=(3.5\pm0.4)\times10^{-3}\,
{\rm Mpc}^{-3}$ (see Table 1 in Cirasuolo et~al.).

Similarly, the SMBH mass function is related to the galaxy SMF as
\begin{eqnarray}\nonumber
\Phi_{\rm G}(z,\mbh)&=&\frac{\rd\log\msph}{\rd \mbh}
\int\rd\log \msph\Phi(z, M_{\rm tot})\\
& &\times\frac{1}{\sqrt{2\pi}\Delta_M}
\exp\left[-\frac{(\log \msph-\log\mbh)^2}{2\Delta_M^2}\right],
\label{nbh_equ}
\end{eqnarray}
where $M_{\rm tot}$ is the total galaxy stellar mass, 
$\msph=M_{\rm tot}({\rm B/T})$ is the mass contained in the bulge, and
$\Phi(z, M_{\rm tot})$ is the galaxy SMF given by P{\'e}rez-Gonz{\'a}lez 
et al. (2008). The factor $\rd\log \msph/\rd\mbh=1/[\ln(10)a_M\mbh]$,
according to Equation (2). 
The galaxy SMF, also expressed as a Schechter function 
(P{\'e}rez-Gonz{\'a}lez et al. 2008), is
\begin{equation}
\Phi(z, M)=\ln(10)\Phi_\star\left(\frac{M}{M_\star}\right)^{1+\alpha}
           \exp\left(-\frac{M}{M_\star}\right).
\label{smf_equ}
\end{equation}
The parameters $\alpha$, $M_\star$, and $\Phi_\star$ are given in Table 2 of 
P{\'e}rez-Gonz{\'a}lez et al. for 12 redshift bins from $z=0$ to $z=4$.

%
\begin{figure}[t]
\centering
\hspace*{0.20cm}\includegraphics[angle=-90.0, width=0.40\textwidth]{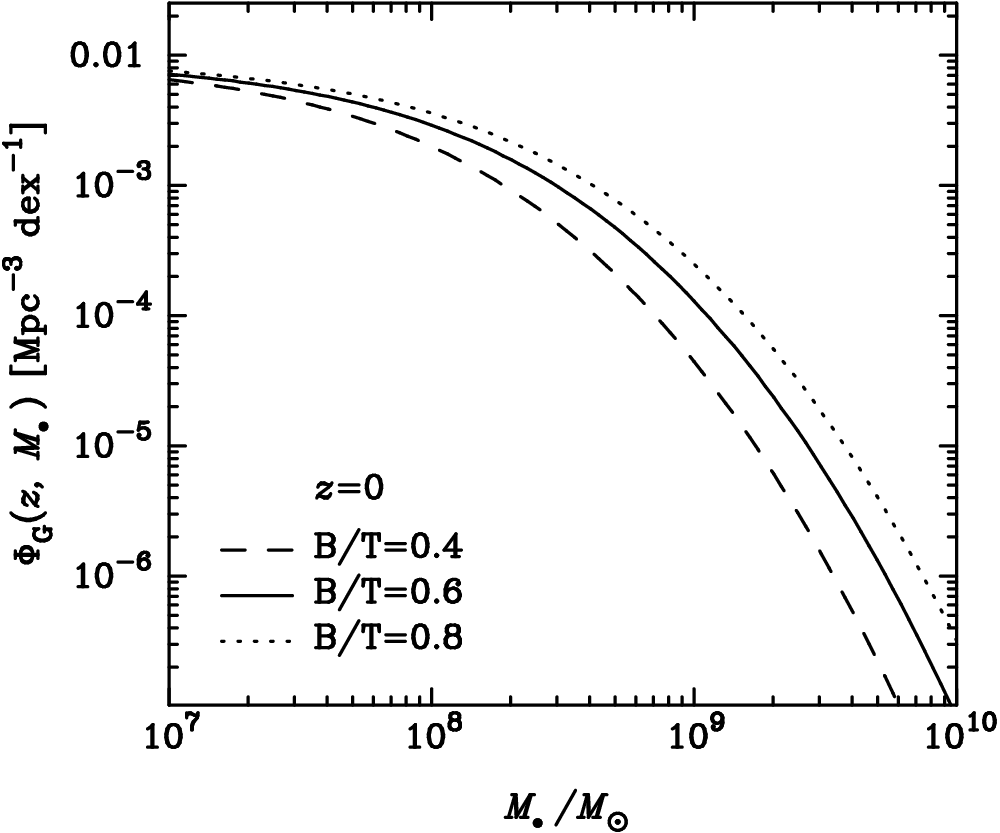}\\
\includegraphics[angle=-90.0, width=0.382\textwidth]{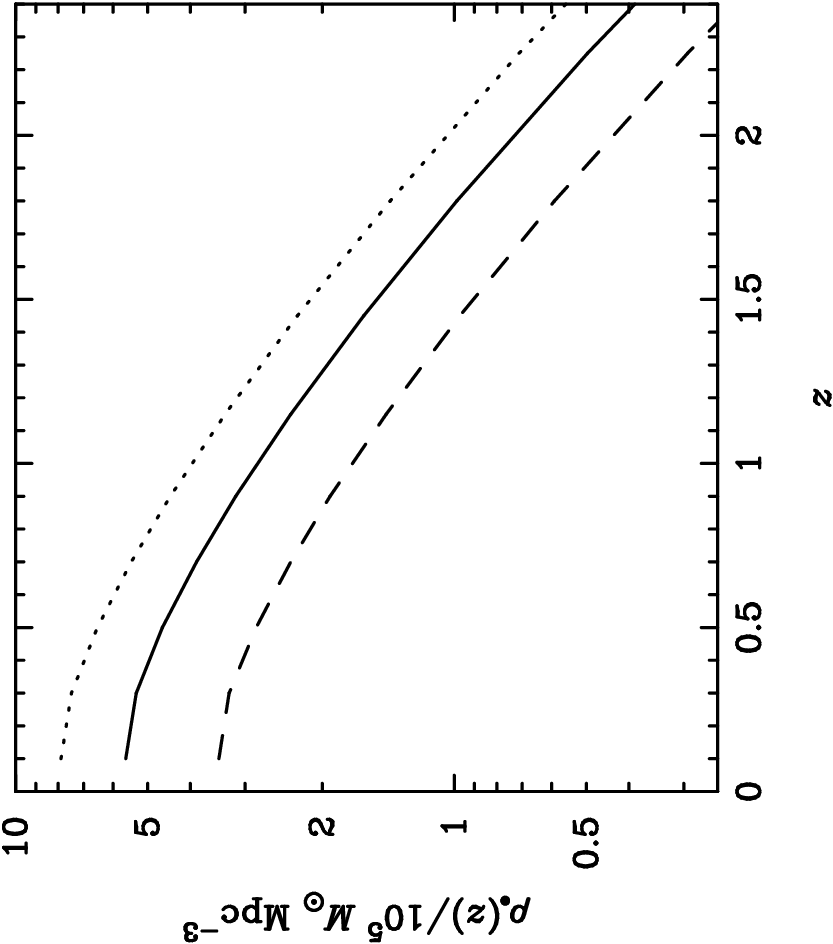}
\caption{SMBH mass function at $z = 0$ ({top}) and mass density as a 
function of $z$ ({bottom}) for B/T = 0.4, 0.6, and 0.8.}
\label{errbt_fig}
\end{figure}
%
%

\section{Results}

\subsection{SMBH Mass Functions}
Figure \ref{nbh_fig} shows the SMBH mass function, computed using the galaxy 
LF and SMF, at $z=0$, 1, and 2.  The lowest redshift bin of Cirasuolo et~al.'s 
LFs is at $z=0.3$, and we extrapolate it to $z=0$.
We confirm that the extrapolation is consistent with the observed
local $K$-band galaxy LF (e.g., \citealt{Devereux09}, and references therein).
The SMBH mass functions 
derived from the galaxy LF and SMF show strikingly good agreement, both in 
shape and in normalization.  A notable discrepancy occurs at $z=2$, where the 
SMF-based SMBH mass function systematically lies above that derived from the 
galaxy LF, for $\mbh$ \gax\ $(2-3) \times 10^8\,M_\odot$.  As this corresponds
roughly to the ``knee" of the mass function, we expect this result to be 
quite sensitive to the values of $M_\star$ and $\Phi_\star$ of the SMF.  
Curiously, inspection of the data in P{\'e}rez-Gonz{\'a}lez et al. (2008) 
reveals that $M_\star$ and $\Phi_\star$ show significant fluctuations precisely 
at the redshift bins $z = 1.8$ and $z = 2.25$; it is unclear whether the 
magnitude of these fluctuations is sufficient to account for deviations in 
the SMF-based SMBH mass function at $z = 2$.

For comparison, we superpose the local ($z \approx 0$) SMBH mass functions 
derived by \cite{Marconi04} and \cite{Vika09}.  The agreement is excellent.
We note that the mass function at the high-mass end hardly evolves 
from $z\approx2$ to $z=0$, indicating that the most massive black holes have 
been largely in place since $z\approx2$ and experience little growth since 
then (\citealt{Marconi04, McLure04}). By contrast, the lower end of the 
mass function undergoes strong evolution.  This trend is often referred to 
as cosmic ``downsizing."

We calculate the mass density for SMBHs for $\mbh\ > 10^7\,M_\odot$ by 
integrating the mass functions and show the results in Figure \ref{nbhz_fig}.  
Once again, the results derived separately from the galaxy LF and SMF are
quite consistent.  The mass density increases rapidly from $z=2$ until 
$z\approx0.5$, below which it begins to saturate, converging toward the local 
($z \approx 0$) mass density of $\rho_\bullet=(4.9\pm0.7)\times10^5\,
M_\odot\,{\rm Mpc}^{-3}$ derived by Vika et al. (2009; black star). 
In light of the accretion paradigm for AGN activity and black hole growth,
this is in line with the redshift evolution of AGN activity, which reaches a 
maximum around $z\approx2-3$ (e.g., \citealt{Ueda03, Hopkins07}) and rapidly 
declines toward lower redshifts, such that by the present-day universe most 
black holes are highly sub-Eddington and nearly quiescent
(\citealt{Ho08, Ho09}).

%
\begin{figure}[t]
\centering
\includegraphics[angle=-90.0, width=0.40\textwidth]{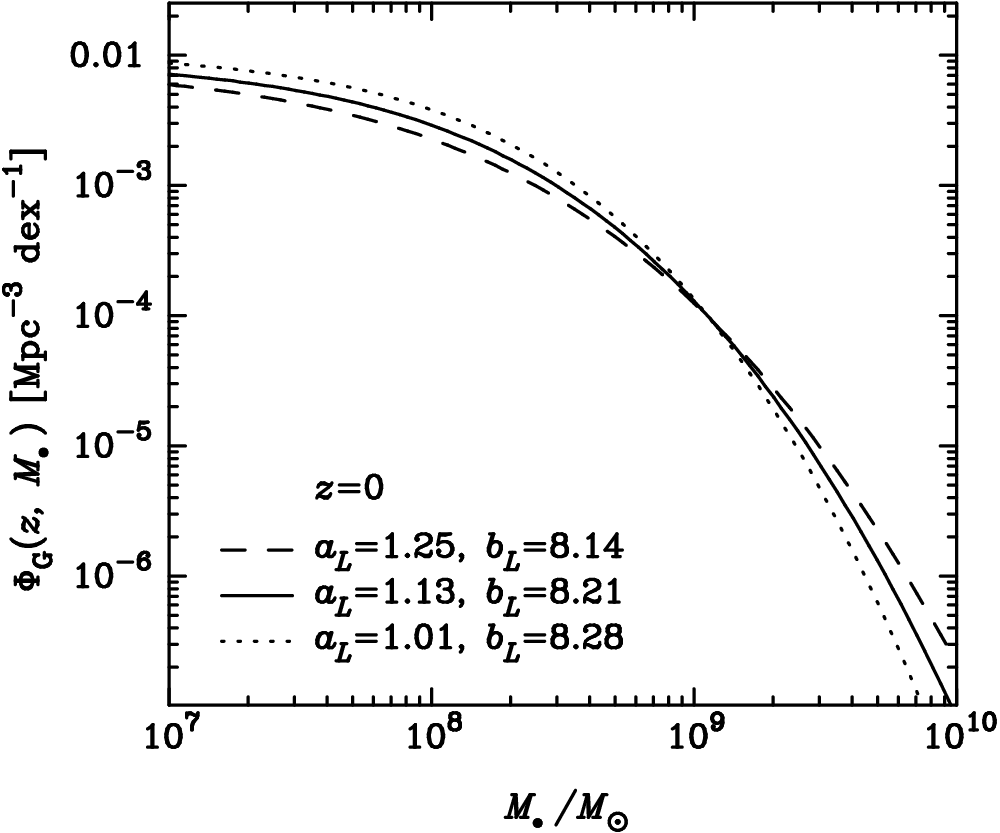}\\
\includegraphics[angle=-90.0, width=0.40\textwidth]{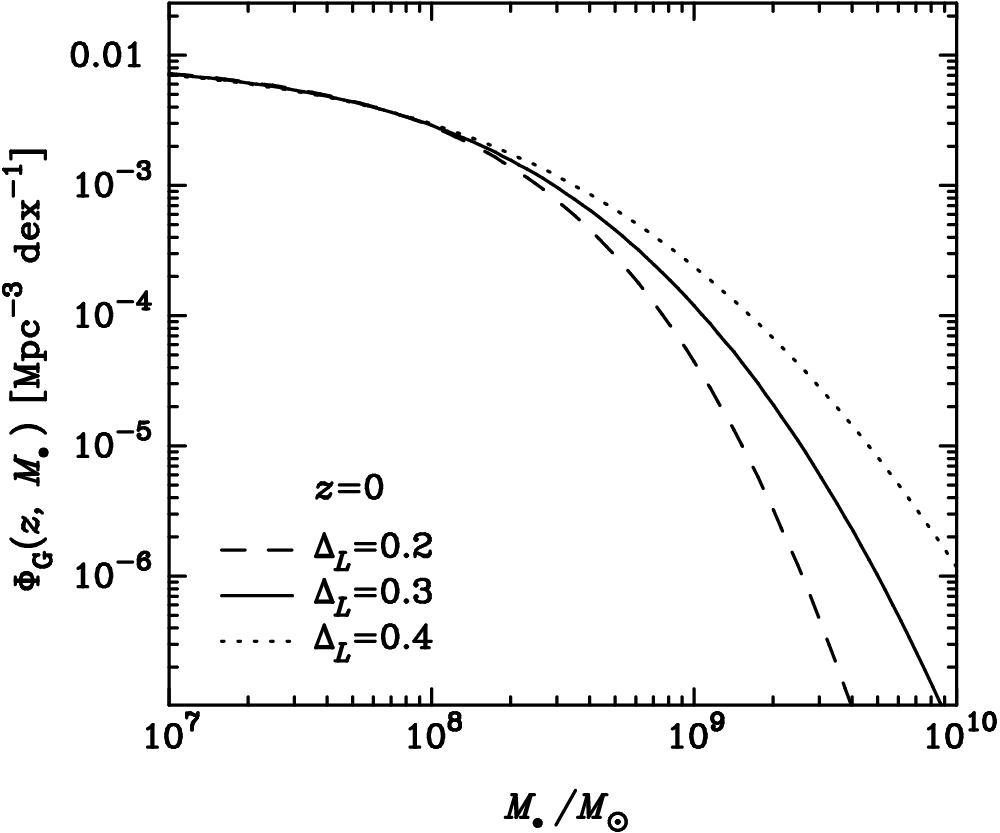}\\
\hspace*{-0.15cm}
\includegraphics[angle=-90.0, width=0.378\textwidth]{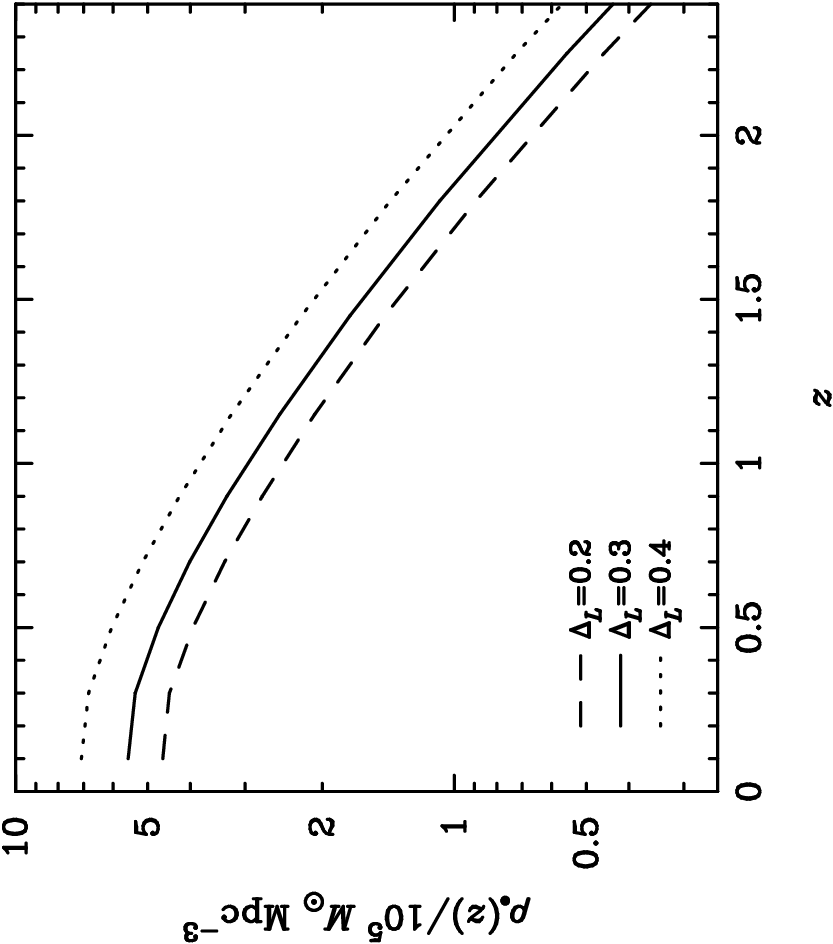}
\caption{Effect on our calculations from the uncertainties from the 
$\mbh-L_{\rm sph}$ relation given by \cite{Marconi03}.  ({Top}) SMBH mass 
functions at $z=0$ for different values of the fitting errors for the 
$\mbh-L_{\rm sph}$ relation.  ({Middle}) SMBH mass functions at $z=0$ for 
three values of the intrinsic scatter $\Delta_L$ for the $\mbh-L_{\rm sph}$ 
relation.  ({Bottom}) SMBH mass densities corresponding to the three mass 
functions of the middle panel.}
\label{errml_fig}
\end{figure}
%
%
%

%
\subsection{Constraints on the Free Parameters}

Table 1 lists the values of the free parameters we used to describe the 
redshift evolution of B/T, $\mbh/\msph$, and luminosity of the stellar 
population.  Their impact on the SMBH mass function can be inferred from 
Equations (\ref{mlz_equ}) and (\ref{mag_equ}). In the Appendix, we explicitly 
show how these parameters affect the SMBH mass density. 

The preceding section explains how we adopt the values of key free parameters 
in our calculations, motivated by various lines of observations. Using the 
galaxy LFs for four bins of morphological types from \cite{Zucca06}, we obtain 
$\gamma=1.3$ for the power-law redshift dependence of B/T and a normalization 
factor of B/T $\approx$ 0.6 at $z=0$. It is interesting to point out that an 
average B/T of 0.6 yields a black hole mass function in good agreement with 
the local observational data, as shown in Figure \ref{nbh_fig}.  Furthermore, 
that the SMBH mass function derived from the galaxy LF matches well those 
obtained from the SMF testifies to the robustness of the spheroid luminosity 
evolution correction adopted in our calculations. Here we additionally show 
that the physical condition that the SMBH mass density always necessarily 
increases over cosmic time imposes additional constraints on the evolution of 
the $\mbh/\msph$ ratio.  In the Appendix, we prove that the SMBH mass density 
derived from the galaxy LF is proportional to  
\begin{equation}
\rho_\bullet(z)\propto(1+z)^{\beta-\gamma a_L}10^{(z/z_M)^{k_M}/2.5-a_L Qz/2.5}\Phi_0(z),
\label{density_equ}
\end{equation}
where $a_L$ is the slope of the $\mbh-\lsph$ relation (see Equation 
(\ref{ml_equ})) and $z_M$, $k_M$, and $\Phi_0(z)$ are the parameters of the 
Cirasuolo et~al. (2010) galaxy LF (see Equations (\ref{mk_equ}) and 
(\ref{phi0_equ})). We note that the parameters $\gamma$ and $\beta$ are 
degenerate along lines of constant $\beta-\gamma a_L$.  Since $a_L\approx1$, 
for the purposes of this discussion let us set $a_L=1$ for convenience. For a 
fixed $\gamma$ and a large value of $\beta$, the first term in Equation 
(\ref{density_equ}) increases monotonically with redshift and dominates over 
the others; an increase of $\rho_\bullet$ with redshift is physically 
prohibited.  Figure \ref{beta_fig} illustrates $\rho_\bullet$ derived for 
three choices of $\beta-\gamma$.  While the case with $\beta-\gamma=-0.4$ 
yields a rapid drop off of $\rho_\bullet$ around $z=0$, the case with 
$\beta-\gamma=0.4$ leads to a negative increase.  
Both cases seem unlikely and conflict with the basic physical requirement 
that $\rho_\bullet$ must increase with time and that most black holes
are nearly quiescent by the present-day universe.  Note that, for $\rho_\bullet$ 
derived from the galaxy SMF, there is a sharp transition between the redshift 
bins $z=0.1$ and $z=0.3$; this may be due to the large uncertainties of the 
SMF at $z=0.1$ (see \citealt{Perez-Gonzalez2008}). Inspection of Figure 
\ref{beta_fig} suggests that, if $\gamma \approx 1.3$, a reasonable value of 
$\beta$ should be $\lesssim1.7$.
Future observations will empirically test this assertion.
%

%
\subsection{Comparison with Previous Results}
In Figure \ref{tamura_fig}, we perform a comparison of our results with those
of \cite{Tamura06}. We find good agreement for the massive end of the mass 
functions, for $\mbh\gtrsim10^{8.3}\,M_\odot$, but the Tamura et al. results 
systematically fall below ours toward lower masses.  We ascribe these 
discrepancies to three factors.

(1) The survey depths of the galaxy LFs are very different. 
\citeauthor{Tamura06} employ the galaxy LFs from the COMBO-17 survey 
(\citealt{Bell04}), whose depth in the $B$ band is $\sim 25$ mag (Vega), which
corresponds to $K \approx 21$ mag (Vega). By comparison, the Cirasuolo et~al. 
LFs reach $K \approx 23$ mag (Vega; \citealt{Lawrence07}), allowing us to 
reach significantly further down the mass function, robustly sampling later
type galaxies and lower mass black holes.

(2) \citeauthor{Tamura06} only consider early-type galaxies, which 
occupy the luminous end of the galaxy LF and host exclusively massive black 
holes.  We, on the other hand, have sufficient sensitivity to sample later 
type galaxies, which have proportionately lower mass bulges and hence 
black holes.  Thus, it is not a surprise that our results agree with those 
of Tamura et al. at the high-mass end of the mass function but diverge 
toward lower masses.  The excellent match between our results and those 
independently obtained by Marconi et al. (2004) and Vika et al. (2009) at 
$z \approx 0$ (Figure \ref{nbh_fig}) testifies to the robustness of our 
assumptions. 

(3) Our calculations take into account the redshift-dependent evolution of B/T 
and $\mbh/\msph$,  while \citeauthor{Tamura06} adopt constant values for 
these parameters. However, for a given total galaxy luminosity and our chosen 
prescription for passive evolution of the spheroid light,\footnote{Like us, 
Tamura et al. (2006) also compute passive evolution using the PEGASE code.} 
B/T and $\mbh/\msph$ have opposite effects on the black hole masses.  In other 
words, the parameters $\gamma$ and $\beta$, which describe the evolution of B/T 
and $\mbh/\msph$, respectively, are degenerate. Because we use $\gamma=1.3$ and 
$\beta=1.4$, in practice these two components nearly cancel each other out in 
terms of their effects on black hole masses at high redshift (see Equations 
(\ref{mlz_equ}) and (\ref{mag_equ})). Therefore, inclusion of the 
redshift-dependent evolution of B/T and $\mbh/\msph$, in fact, induces only 
mild differences.

%
\section{Sources of Uncertainties}

As the galaxy LF and SMF are well described by a Schechter function,
the exponential decline beyond the knee implies that even minor changes in the
high-luminosity or high-mass end can induce significant effects on the upper 
end of the SMBH function.  Here we 
examine the impact of various sources of uncertainties on our derivation of 
the SMBH mass function.  For simplicity, we report only on results based on 
the galaxy LF; the results based on the galaxy SMF are similar.

%
\begin{figure}[t]
\centering
~~~~\includegraphics[angle=-90.0, width=0.40\textwidth]{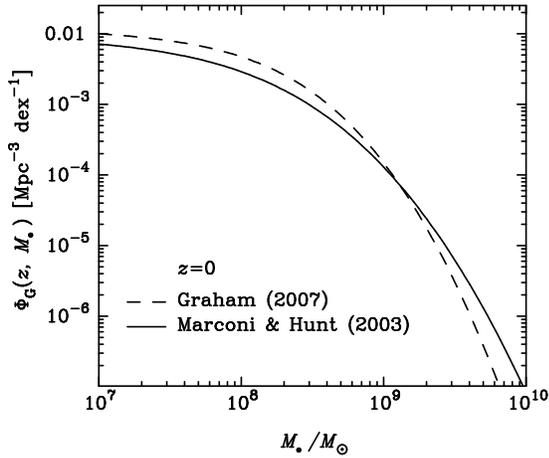}
\caption{SMBH mass function at $z=0$ for the $\mbh-L_{\rm sph}$ relation from 
Marconi \& Hunt (2003; solid line), which is assumed in our calculations, and 
for that from Graham (2007; dashed line), who reanalyzed the data of 
Marconi \& Hunt.}
\label{errgraham_fig}
\end{figure}
%
%
%
%
\begin{figure}[t]
\centering
\includegraphics[angle=-90.0, width=0.38\textwidth]{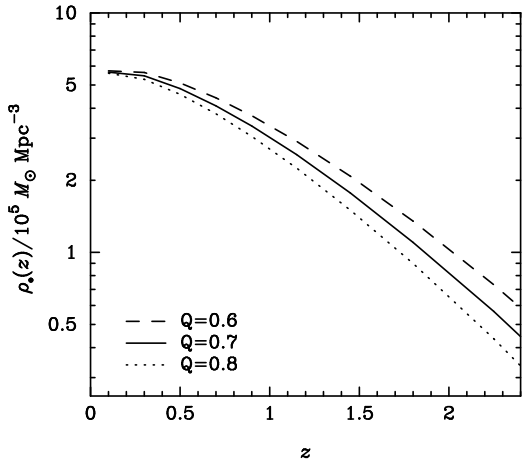}
\caption{SMBH mass density for three choices of the passive evolution 
parameter, $Q$ = 0.6, 0.7, and 0.8.}
\label{errpass_fig}
\end{figure}
%
%
\subsection{Bulge-to-total Luminosity Ratio}
Figure \ref{errbt_fig} shows SMBH mass functions and mass densities for 
B/T = 0.4, 0.6, and 0.8 at $z=0$. As to be expected because of the exponential 
drop off of the galaxy LF, the high-mass end of the black hole mass function 
is especially sensitive to the value of B/T.  
For a given total galaxy luminosity, larger B/T leads to larger black hole 
masses and hence to a larger number density and integrated mass density of 
SMBHs. 
The parameter $\gamma$, which measures the redshift evolution of B/T, is 
degenerate with respect to $\beta$. Therefore, we treat $\beta-\gamma a_L$ as 
one independent parameter and have shown its influence on the SMBH mass 
density in Section 3.2 and in Figure \ref{beta_fig}.

In our calculations, for the sake of simplicity, we neglect the dependence of 
B/T on the galaxy luminosity or mass, and just focus on its average 
dependence on redshift.  \cite{Schechter87} show that the bulge-to-disk ratio 
varies weakly with galaxy luminosity, roughly as $(L/L_\star)^{0.28}$ (see 
also Figure 12 of \citealt{Benson07}).  For $L \approx 0.1L_\star$ to 
$10 L_\star$, B/T increases from $\sim$0.45 to 0.75, if we normalize B/T to 
0.6 for $L=L_\star$.  This luminosity-dependent variation of B/T is 
small, and we neglect it.
We further neglect any intrinsic dispersion in the average value of B/T at a 
given morphological type, as this dispersion is poorly quantified even 
for nearby galaxies (e.g., Simien \& de~Vaucouleurs 1986; de~Jong 1996). 

%
\subsection{The Local $\mbh-L_{\rm sph}$ Relation}
The upper panel of Figure \ref{errml_fig} shows the fitting errors of the 
$\mbh-L_{\rm sph}$ relation derived by \cite{Marconi03} and their 
corresponding influences on our results. We find that the reported 
uncertainties on the slope and normalization ($a_L$ and $b_L$) of the 
$\mbh-L_{\rm sph}$ relation induce only minor effects on the overall SMBH mass 
function. \cite{Marconi03} reported an intrinsic scatter of $\Delta_L 
\approx0.3$ for the $\mbh-L_{\rm sph}$ relation.  To assess its impact on 
our results, we calculate the mass function and mass density for three 
values of $\Delta_L$ (0.2, 0.3, and 0.4), as shown in the middle and bottom 
panels of Figure \ref{errml_fig}.  We find that $\Delta_L$ has a remarkably 
large impact on the shape of the upper end of the mass function.  This is again due 
to the exponential decline of the galaxy LF.  However, the effect of 
$\Delta_L$ on the mass density is relatively insignificant, changing 
the overall amplitude by $\sim \pm 25\%$.

\cite{Graham07} reanalyzed the work of \cite{Marconi03}, updating some of the 
data on black hole masses and galaxy magnitudes.  He reports a somewhat 
shallower $\mbh-L_{\rm sph}$ relation, with best-fit parameters 
$(a_L, b_L, \Delta_L)=(0.93, -1.80, 0.33)$.  Figure \ref{errgraham_fig} 
compares the effect that the two versions of the $\mbh-L_{\rm sph}$ relation 
have on our derived SMBH mass function, assuming an intrinsic scatter of 
$\Delta_L=0.3$. The differences are minor over all masses.

Two caveats should be kept in mind.  First, recent studies suggest that 
classical bulges and pseudo-bulges may obey somewhat different scaling relations with black hole 
mass (e.g., Greene et al. 2008; Jiang et al. 2011; 
Kormendy et al. 2011; Xiao et al. 2011).  Another complication arises from the 
possibility that, at a given redshift, galaxies in different evolutionary 
phases may have different values of $\mbh/\msph$ (e.g., \citealt{Alexander08, 
Lamastra10}).  Unfortunately, the current data on the bulge properties of 
high-redshift galaxies are far too crude to allow us to concretely address 
either of these issues.

\subsection{Choice of Passive Evolution}

We assume that the stellar population of the bulge formed in a single burst 
at a formation redshift of $z_f$=3, 4, and 10, thereafter fading passively 
as $M_K(z)=M_K(z=0)-Qz$, with $Q\approx0.7$.  As we can see from Figure 
\ref{passive_fig}, the value of $Q$ is quite insensitive to $z_f$. In 
Figure \ref{errpass_fig}, we examine the sensitivity of the SMBH mass density 
to the value of $Q$, choosing values of 0.6, 0.7, and 0.8.  Variations of 
this magnitude on $Q$ have only a mild influence on $\rho_\bullet$.

%
%
\subsection{Summary of the Uncertainties}

Among the uncertainties discussed above, the most important are those related 
with the local $\mbh-L_{\rm sph}$ relation.  While uncertainties in the slope 
and zero point of the relation affect the SMBH mass function at the level of
\lax\ 0.2 dex over the entire mass range, the upper end of the mass function
is especially sensitive to the intrinsic scatter of the $\mbh-L_{\rm sph}$ 
relation.  Changing the scatter at the level of $\pm 0.1$ dex affects 
the mass function by $\sim \pm 0.5$ dex at $\mbh\ \approx 2\times10^9\,M_\odot$.
This is due to the exponential falloff of the galaxy LF, such that even a 
minor change in luminosity leads to a significant variation in the galaxy 
number density, and hence in the SMBH number density.  In a similar manner,
the upper end of the SMBH mass function is sensitive to the values of B/T. In 
our calculations, the redshift evolution of B/T, encapsulated in the parameter
$\gamma$, is degenerate with the evolution of $\mbh/\msph$, which is 
parameterized by $\beta$.  Fortunately, for reasonable choices of $\gamma$ and 
$\beta$, which are well motivated by current observations, the redshift 
evolution of B/T and $\mbh/\msph$ largely cancel out.  Plausible variations
on the prescription for passive evolution of the spheroid luminosity also 
introduce relatively minor perturbations on our results.  In summary, from 
inspection of the tests presented in Figures \ref{beta_fig} and 
\ref{errbt_fig}--\ref{errpass_fig}, it seems that the total 
uncertainties on the SMBH mass function are generally within $\sim$0.3 dex.

Lastly, it is worth emphasizing that the stellar masses used to compute the 
galaxy SMF of \cite{Perez-Gonzalez2008} assume a Salpeter IMF.  As has been 
pointed out by a number of authors (e.g., \citealt{Bell03, Bruzual03, 
Pozzetti07}), the Salpeter IMF is too rich in low-mass stars and might 
overestimate the stellar masses.  The \cite{Chabrier03} IMF may yield more 
reasonable stellar masses (e.g., \citealt{di_Serego05}).  Stellar masses 
derived with the Chabrier IMF are smaller than those derived with the
Salpeter IMF by a factor of $\sim1.7$ (e.g., \citealt{Pozzetti07}). 
Consequently, the characteristic mass $M_\star$ of the 
\citeauthor{Perez-Gonzalez2008} galaxy SMF would be lower by this amount.
In the power-law regime ($M\lesssim M_\star$), this is equivalent to a 
decrease of $\Phi_\star$ by a factor of $\sim(1.7)^{1+\alpha}$ (see Equation 
(\ref{smf_equ})). For a characteristic value of $\alpha\approx -1.2$ 
(P{\'e}rez-Gonz{\'a}lez et al. 2008), $\Phi_\star$, and thus the magnitude of 
SMBH mass function, decreases slightly by $\sim0.05$ dex in the power-law 
regime.  Beyond the knee ($M>M_\star$), however, the exponential decline of 
the SMF significantly affects the magnitude of SMBH mass function.
Nevertheless, the resultant SMBH mass density decreases
systematically only by a factor of $\sim1.7$ (0.23 dex).

However, the existing observations of the local SMBH mass function
and mass density (e.g., Marconi et al. 2004; Vika et al. 2009) place
constraints on the adopted IMF parameters. If the Chabrier 
IMF is adopted, a mismatch will be induced between the SMBH mass function and
the existing observations, and other parameters in the calculations will have
to be modified accordingly (e.g., increasing B/T). Note that if the
Chabrier IMF is universal and invariant with redshift, it will not affect
the overall evolution of the SMBH mass function, and the present results
remain unchanged.

%
\section{Summary}
We derive the SMBH mass function from $z = 0$ to $z = 2$ using the up-to-date,
deep, wide-area $K$-band galaxy LF of Cirasuolo et~al. (2010), and, in a 
complementary manner, using the galaxy SMF of P{\'e}rez-Gonz{\'a}lez et al. 
(2008).  In addition to extending much further down the mass function than 
previous studies, enabling us to robustly sample later type galaxies and 
lower mass black holes, our analysis carefully considers redshift-dependent 
corrections to the average B/T of the galaxy populations, the 
$\mbh/\msph$ ratio, and the stellar luminosity of the bulge.  We find 
excellent agreement between the SMBH mass functions derived from the galaxy 
LFs and the galaxy SMFs.  Moreover, the resultant SMBH mass function and 
integrated mass density for the local epoch match well those derived 
independently by other studies.

In a companion paper, we will use the SMBH mass function obtained in this work 
to determine the radiative efficiency of black hole accretion as a function of
redshift, with the goal of constraining the cosmological evolution of black hole spin.

%
\acknowledgements{Y.R.L. thanks N. Tamura for providing the data of their SMBH 
mass function, and F. Shankar and X.-W. Cao for useful discussions. We thank 
the referee for helpful suggestions, and members of the IHEP AGN group for 
discussions. This research is supported by NSFC-10733010, -10821061 and -11173023, and 973 
project (2009CB824800).  The work of LCH is supported by the Carnegie 
Institution for Science.}

\appendix
\section{Redshift dependence of SMBH mass density}
In this appendix, we clarify how the SMBH mass density depends on certain 
free parameters adopted in our analysis.  The Schechter (1976) function used 
to describe the galaxy LF has the general form 
\begin{equation}
\Phi(z, L)\rd L=\Phi_0(z)\left(\frac{L}{L_\star}\right)^{-\alpha}
                \exp\left(-\frac{L}{L_\star}\right)\frac{\rd L}{L_\star}, 
\end{equation}
where $\Phi_0(z)$ is the normalization and $L_\star$ is the characteristic 
luminosity as a function of redshift. If, for the sake of simplicity, we 
neglect the intrinsic scatter of the $\mbh-L_{\rm sph}$ relation in Equation 
(\ref{ml_equ}), the SMBH mass function $\Phi_{\rm G}(z, \mbh)$ is related 
to the galaxy LF as
\begin{equation}
\Phi_{\rm G}(z, \mbh)\rd\mbh=\Phi(z, L)\rd L.
\end{equation}
Integrating the SMBH mass function, we obtain the total black hole mass 
density 
\begin{equation}
\rho_\bullet(z)=\int\Phi_{\rm G}(z, \mbh)\mbh\rd\mbh 
         \propto\int\Phi(z, L)\mbh\rd L.
\end{equation}
According to the $\mbh-L_{\rm sph}$ relation (Equation 7), the black hole mass 
can be expressed as
\begin{equation}
\mbh=10^{b_L-a_L Qz/2.5}(1+z)^\beta [({\rm B/T})L]^{a_L}.
\end{equation}
Hence,
\begin{equation}
\rho_\bullet(z)=10^{b_L-a_L Qz/2.5}(1+z)^\beta\Phi_0(z)L_\star(z)({\rm B/T})^{a_L}
                \Gamma(1-\alpha+a_L),
\end{equation}
where $a_L$ is the slope of the $\mbh-\lsph$ relation (see Equation 
(\ref{ml_equ})), $\Phi_0(z)$ is given by Equation (\ref{phi0_equ}), and 
$\Gamma(x)$ is the Gamma function. Since the characteristic luminosity from 
Equation (\ref{mk_equ}) is
\begin{equation}
 L_\star(z)=L_\star(z=0)10^{(z/z_M)^{k_M}/2.5},
\end{equation}
we have
\begin{equation}
\rho_\bullet(z)\propto(1+z)^{\beta-\gamma a_L}10^{(z/z_M)^{k_M}/2.5-a_L Qz/2.5}\Phi_0(z).
\end{equation}
We note that, apart from being sensitive to the details of the galaxy LF, the 
redshift evolution of the SMBH mass density depends on the chosen prescription 
for the redshift evolution of B/T, the $\mbh-L_{\rm sph}$ relation, and 
luminosity variation of the stellar population. The parameters $\gamma$ and 
$\beta$ are degenerate along lines of constant $\beta-\gamma a_L$. The physical 
requirement that $\rho_\bullet$ always increases with time places additional 
constraints on the redshift dependence of these parameters (see also 
\citealt{Hopkins06}).


\end{document}